# Energetics of thermal inactivation of excitonically-induced defect formation in rare-gas solids

*A.N. Ogurtsov, N.Yu. Masalitina, O.N. Bliznjuk*

*National Technical University "KhPI", Frunse Street 21, Kharkov 61002, Ukraine*

Selective excitation of excitons in rare-gas cryocrystals by photons with energies $h\nu < E_g$ results in accumulation of Frenkel-pairs by intrinsic excited-state mechanism of defect formation *via* self-trapping of excitons [1]. Recently the simple kinetic model was proposed, which allows fitting the experimental dose dependences of "defect" subbands and obtaining the particular kinetic parameters [2]. Application of this model provided a way of qualitative and quantitative analysis and certification of rare-gas crystals, which is indispensable at any attempt of comparison of data from different samples. At the same time it is well known that there is a strong thermal quenching of the defect formation processes, which was initially explained by temperature dependence of lifetime of emitting states [3]. Moreover, in the same temperature range the electron traps become active and charge recombination processes result in rich spectra of thermoluminescence [4]. But even without discussion of the detailed microscopic nature of such inactivation processes we can consider the general case of the reversible transition between active and inactive exciton-trapping states with equilibrium constant $K = \exp[(T\Delta S - \Delta H)/kT]$, where $\Delta H$ and $\Delta S$ – are the enthalpy and entropy of inactivation.

In the present study we collect last years dose measurements of luminescence spectra evolution under selective synchrotron irradiation and apply the Eyring's transition state concept [5] to the processes of thermal activation-inactivation of exciton trapping states. The experiments were carried out at the SUPERLUMI-station at HASYLAB, DESY, Hamburg. The selective photon excitation was performed with spectral resolution $\Delta\lambda = 0.2$ nm. The VUV-luminescence analysis was performed with $\Delta\lambda = 2$ nm, Pouey high-flux monochromator equipped with a multisphere plate detector. The dose curves at different temperatures under irradiation by photons with energies $E < E_g$ were measured. These curves are saturated at long time of irradiation therefore we used the slopes of the initial linear parts of the dose curves at $t = 0$ [3] as the defect formation rates $w$ (Fig.1).

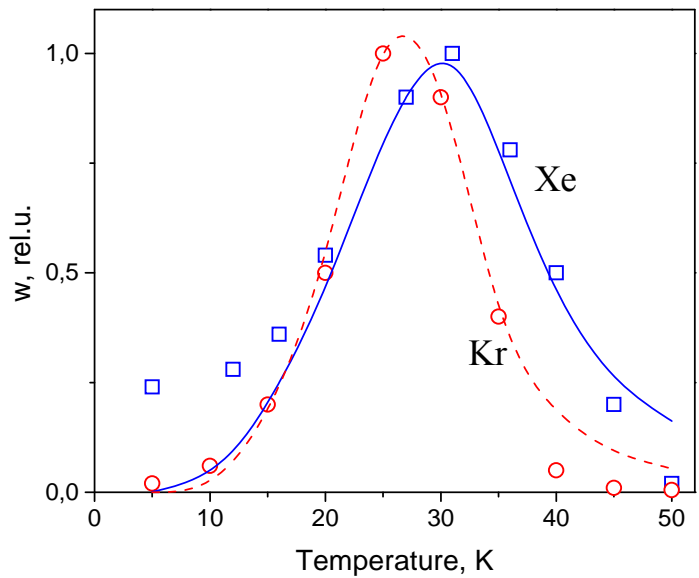

Figure 1: Temperature dependences of defect accumulation rate of solid Xe (□) and Kr (○); and their fitting by Eq. 1 (solid and dashed curves).

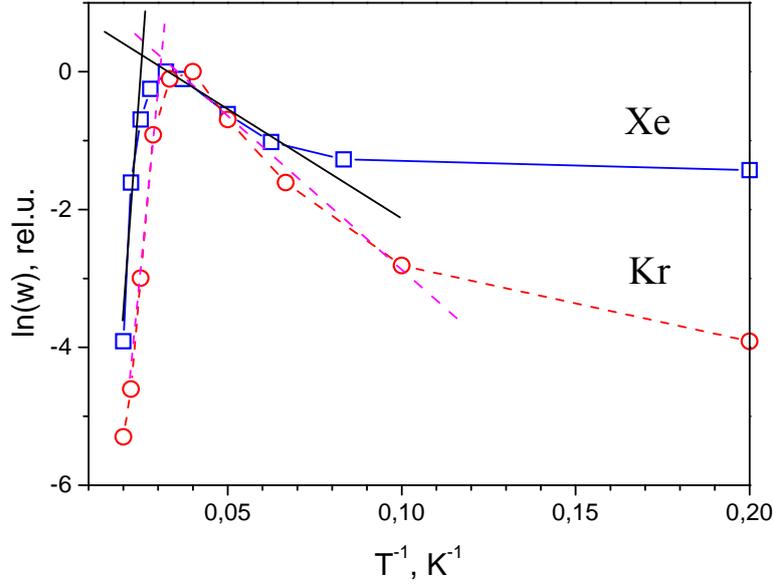

Figure 2: Semilogarithmic plot of defect accumulation rates of solid Xe (□) and Kr (○) as a function of reciprocal temperature, and their linear asymptotics (straight lines).

Figure 1 shows measured defect formation rates (points) and calculated curves for solid Xe and Kr. Following Eyring assumption we can fit the temperature dependence of the defect formation rate as

$$w(T) = \beta \cdot \frac{T \cdot \exp\left(-\dfrac{E}{kT}\right)}{1 + \exp\left(\dfrac{\Delta S}{k}\right) \cdot \exp\left(-\dfrac{\Delta H}{kT}\right)}, \quad (1)$$

where $k$ – the Boltzmann's constant, $E$ – Arrhenius activation energy, $\beta$ – scaling factor. The values of activation energy $E$ and enthalpy of inactivation $\Delta H$ may be determined from the upper and lower tangents of the $w(T)$ in the coordinates ($\ln[w(T)]$) vs. ($T^{-1}$) (Fig. 2). At high $1/T$ the temperature dependence of the defect formation rate tends to $\ln(w(T)) = (-E/k)(T^{-1})$. At high temperatures ($T^{-1} \to 0$) the temperature dependence tends to $\ln(w(T)) = [-(\Delta H - E)/k)](T^{-1})$. The value of entropy of inactivation $\Delta S$ may be obtained from the equilibration condition $d(\ln[w(T_{max})])/dT = 0$, which can be expressed in the form

$$\frac{E + kT_{max}}{\Delta H - E - kT_{max}} = K = \exp\left(-\frac{\Delta H}{kT}\right) \cdot \exp\left(\frac{\Delta S}{k}\right), \quad (2)$$

where $T_{max}$ – the position of the maximum of $w(T)$. The best fit of the data is obtained with $T_{max}^{Xe} = 30$ K, $T_{max}^{Kr} = 27$ K; $E_{Xe} = 2.8$ meV, $E_{Kr} = 4$ meV; $\Delta H_{Xe} = 28$ meV, $\Delta H_{Kr} = 30$ meV; $\Delta S_{Xe} = 0.8$ meV·K$^{-1}$, $\Delta S_{Kr} = 1$ meV·K$^{-1}$. The low-temperature misfitting originates from free exciton diffusion contribution [3]. The evaluation of the energetic parameters of inactivation processes for the case of solid Ar and Ne is in progress.